\documentclass[aps,prb,10pt,twocolumn,showpacs,amsmath,amssymb,superscriptaddress,floatfix]{revtex4-2}
\usepackage{graphicx}% Include figure files
\usepackage{dcolumn}% Align table columns on decimal point
\usepackage{bm}% bold math
\usepackage{hyperref}% add hypertext capabilities
\hypersetup{colorlinks,citecolor=blue, filecolor=blue, linkcolor=blue , urlcolor=blue}
\usepackage[utf8]{inputenc}
\usepackage{newunicodechar}
\newunicodechar{≈}{\approx}
\newunicodechar{⁻}{$^{-}$}
\newunicodechar{₀}{$_0$}

\begin{document}

\title{Temperature Evolution of Magnon-Polarons in the van der Waals Antiferromagnet FePS\texorpdfstring{$_3$}{3}}

\author{A.~Krajewska}
        \email{aleksandra.krajewska@pw.edu.pl}
        \affiliation{CENTERA, CEZAMAT, Warsaw University of Technology, 02-822 Warsaw, Poland} 
        \affiliation{CENTERA Labs, Institute of High Pressure Physics, PAS, 01-142 Warsaw, Poland}

\author{J.~Kierdaszuk}
        \affiliation{CENTERA, CEZAMAT, Warsaw University of Technology, 02-822 Warsaw, Poland}

\author{M.~Dub}
    \affiliation{CENTERA, CEZAMAT, Warsaw University of Technology, 02-822 Warsaw, Poland}
    \affiliation{CENTERA Labs, Institute of High Pressure Physics, PAS, 01-142 Warsaw, Poland}

\author{E.~Karpierz-Marczewska}
    \affiliation{CENTERA, CEZAMAT, Warsaw University of Technology, 02-822 Warsaw, Poland}
    \affiliation{CENTERA Labs, Institute of High Pressure Physics, PAS, 01-142 Warsaw, Poland}

\author{M.~Potemski}
    \email{marek.potemski@lncmi.cnrs.fr}
    \affiliation{CENTERA, CEZAMAT, Warsaw University of Technology, 02-822 Warsaw, Poland}
    \affiliation{Laboratoire National des Champs Magn\'etiques Intenses, LNCMI-EMFL, CNRS UPR3228, Univ. Grenoble Alpes, Univ. Toulouse, Univ. Toulouse 3, INSA-T, Grenoble and Toulouse, France}
    \affiliation{Faculty of Physics, University of Warsaw, 02-093 Warsaw, Poland}

\date{\today}

\begin{abstract}
Magnon polarons, hybrid excitations arising from magnon–phonon coupling, were investigated in the van der Waals antiferromagnet FePS$_3$ using magneto-Raman scattering measurements at temperatures ranging from $4$~K to the vicinity of the Néel temperature (T$_N=118$~K). The lower Zeeman-split component of the magnon-like mode, tuned by the magnetic field toward the lower-energy phonon mode, exhibits an avoided crossing. The corresponding magnetic field decreases from approximately $13$~T at $4$~K to $5.6$~T at $85$~K, bringing the effect within reach of a standard laboratory magnet. A model of three coupled bosonic modes shows that the magnon–phonon coupling constant, approximately $2.7$~cm$^{-1}$ at low temperatures, decreases significantly with increasing temperature, following a trend similar to the temperature-induced softening of the magnon mode. This decrease in magnon–phonon coupling is discussed in terms of a possible reduction in sublattice magnetization due to thermal fluctuations.
\end{abstract}

\maketitle 
\section{INTRODUCTION}

Antiferromagnetic van der Waals materials belonging to the MPX$_3$ family~\cite{Grasso2002, Jiang}, where M = Fe, Cr, Co, Mn, or Ni and X = S or Se, have attracted considerable attention as promising platforms for studying two-dimensional magnetism and developing miniaturized spintronic and magnonic devices \cite{Jiang, Burch2018, Gibertini2019}. Coupling between spin-wave excitations (magnons) and lattice vibrations (phonons) is a phenomenon of particular interest in these materials \cite{StreibPRB19, BozhkoLTP20, ManPRB17, DreherPRB12}. Hybridization between magnon and phonon modes can modify excitation lifetimes, energy relaxation, and transport, with important implications for thermal and magnetic dynamics in antiferromagnets \cite{KikkawaPRL16, Ilyas2024, Luo2025}. Hybrid magnon--phonon modes (magnon--polarons) in MPX$_3$ materials can be investigated using optical spectroscopy, particularly Raman scattering \cite{McCrearyPRB20, Liu2021, PhysRevB.104.134437, doi:10.1126/sciadv.abj3106}. Although these techniques primarily probe excitations near zero wavevector, magnon--phonon hybridization can be identified by observing the repulsion between magnon and phonon modes as an external parameter, such as a magnetic field, hydrostatic pressure and temperature tune their relative energies. Magnon--polarons have been reported in several members of the MPX$_3$ family~\cite{doi:10.1126/sciadv.abj3106, Luo2023NL, Cui2023, Jana, JanaCoPS3}. Among these materials, FePS$_3$ has attracted particular attention~\cite{Liu2021, PhysRevB.104.134437,zhang2021coherentstrongcouplingterahertzmagnons, Pawbake} because its magnon gap lies in the terahertz frequency range, close to the energies of several optical phonons. In uniaxial antiferromagnets such as FePS$_3$, an external magnetic field provides a natural means of tuning the magnon and phonon modes into resonance and thereby quantifying their coupling. The magnetic field splits the magnon mode into two Zeeman components~\cite{McCrearyPRB20, Liu2021, PhysRevB.104.134437, KefferKittel1952, Wyzula}, which can be tuned across phonon modes. The resulting avoided crossings provide a direct measure of the magnon--phonon coupling strength.

Within a spin-wave description, the effective linear magnon–phonon coupling is determined by the microscopic spin–lattice interactions and the magnetic ground state~\cite{StreibPRB19, Vaclavkova_2020, Wang2023PCCP}. Consequently, the amplitude of the effective coupling strength might depend on the ordered magnetic moment.
Investigating this dependence in FePS$_3$ requires a control parameter that modifies the magnetic order. An external magnetic field changes the magnon--phonon detuning through the Zeeman effect, allowing the coupling strength to be determined from the resulting avoided crossing \cite{Liu2021, PhysRevB.104.134437, Pawbake}. However, the magnetic field leaves the magnetic order largely unchanged, particularly in FePS$_3$, with the spin-flop transition estimated in the range of $E_M^0/(g\mu_\mathrm{B}) \approx 124$~T. Increasing the temperature, by contrast, provides a natural means of weakening of the magnetic order due to progressive appearance of thermal fluctuations. This reduction is expected to be accompanied by a softening of the magnon mode, while the phonon energies are perceived to remain comparatively weakly affected up to the N\'eel temperature, $T_\mathrm{N} \approx 118$~K~\cite{McCrearyPRB20, PhysRevB.104.134437, LeeNL16}. Magnon-polarons in FePS$_3$ have so far been studied predominantly at the lowest temperatures, where the magnetic order is essentially saturated~\cite{Liu2021, PhysRevB.104.134437, zhang2021coherentstrongcouplingterahertzmagnons}.

In this work, we investigate how magnon--phonon hybridization in FePS$_3$ evolves with increasing temperature, focusing in particular on whether the reduction in magnetic order modifies the magnon--phonon coupling strength. We study bulk FePS$_3$ crystals using magneto-Raman spectroscopy in magnetic fields up to 9~T at fixed temperatures ranging from 4 to $125$~K. Our analysis focuses primarily on the coupling between the lower Zeeman-split magnon branch, denoted $M_-$, and a nearby phonon mode, denoted $P_3$.

For temperatures up to $85$~K, we can reliably track the magnon-like Zeeman-split branches $M_-$ and $M_+$, together with the phonon-like mode $P_3$, as functions of the applied magnetic field. Their behavior is described using a model of three coupled bosonic modes, $M_-$, $M_+$, and $P_3$~\cite{PhysRevB.104.134437, Pawbake}. We assume that both magnon branches couple to the phonon mode with the same coupling strength, $\lambda$. By fitting the model to the experimental data at each temperature, we extract the bare magnon and phonon energies, the effective $g$-factor, and the magnon--phonon coupling strength. The same model also reproduces the splitting of the magnon-like mode at zero magnetic field and its evolution with temperature. We discuss the temperature dependence of the extracted parameters in the context of the effects which can be driven by weakening of the magnetic order.

\section{MATERIAL AND EXPERIMENTAL DETAILS}

\begin{figure}[t]
    \includegraphics[width=0.8\columnwidth]{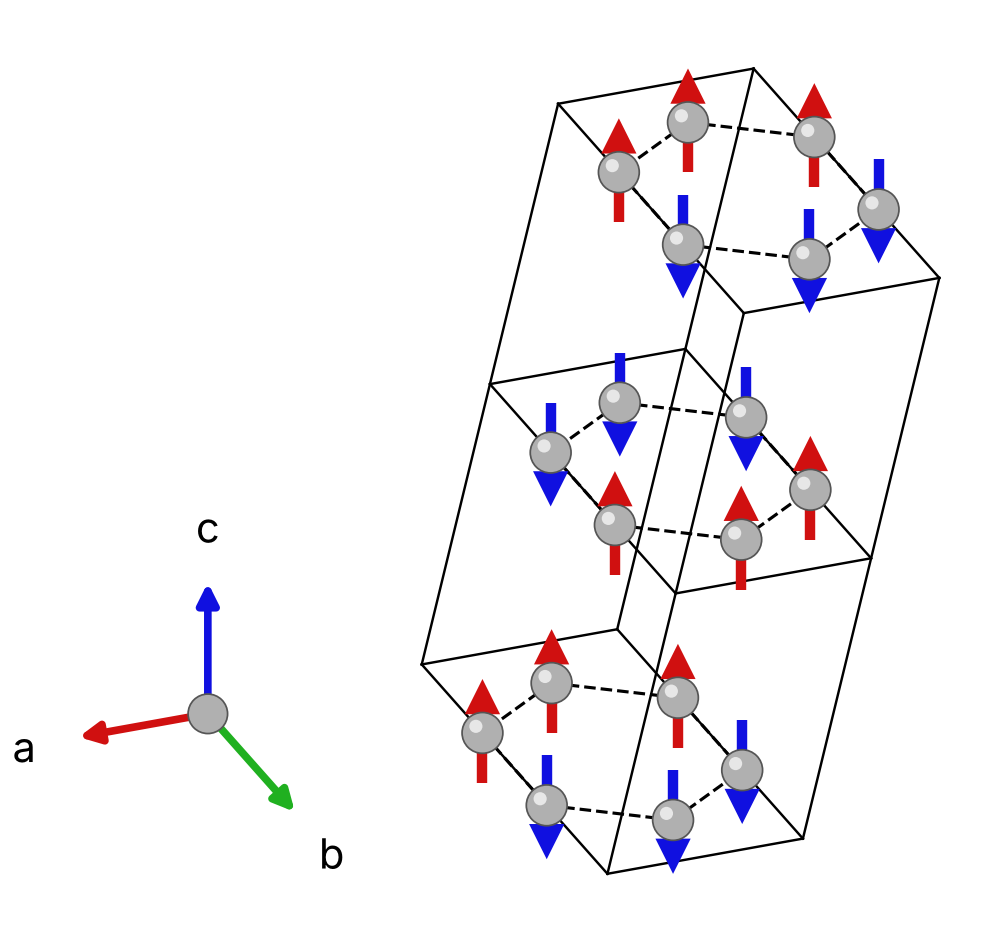}
    \caption{\label{Fig1} Schematic view of the magnetic structure of FePS$_3$ in the antiferromagnetic phase. Only the Fe$^{2+}$ ions (spheres) are shown for three consecutive layers; dashed lines mark the Fe honeycomb lattice and solid lines the monoclinic unit cell. Red and blue arrows denote the Fe$^{2+}$ spins pointing up and down, respectively, perpendicular to the layer planes. Within each layer, ferromagnetic zigzag chains are coupled antiferromagnetically to each other, and the spin orientation is reversed between neighboring layers. The atomic positions follow Ref.~\cite{LanconPRB16}.}
\end{figure}

FePS$_3$ is a layered van der Waals antiferromagnet that crystallizes in a monoclinic structure with space group C2/m \cite{McCrearyPRB20, LeeNL16} and lattice parameters $a = 0.595$~nm, $b = 1.030$~nm, $c = 0.672$~nm, $\beta = 107.16^\circ$ \cite{OuvrardMRB85}. Within each layer, the Fe$^{2+}$ ions ($S = 2$) form a honeycomb lattice in the $ab$ plane and are surrounded by sulfur atoms in octahedral coordination. The layers are held together by weak van der Waals interaction. Below the Néel temperature $T_N \approx 118$ K, the Fe spins order in a zigzag antiferromagnetic structure, in which ferromagnetic zigzag chains are coupled antiferromagnetically to the neighboring chains and the spins are aligned perpendicular to the layer planes, while the exchange between the layers is very small \cite{LanconPRB16} (see Fig.~\ref{Fig1}).

The experiments were performed on commercially available synthetic FePS$_3$ single crystals, which, according to the supplier, were characterized by X-ray diffraction and energy-dispersive X-ray spectroscopy. Thin flakes, a few micrometers thick, were exfoliated from the bulk crystals and deposited on Si/SiO$_2$ substrates.

Magneto-Raman scattering experiments were carried out in the Faraday geometry, with the magnetic field applied perpendicular to the layer planes, i.e., parallel to the direction of light propagation, at magnetic fields up to $9$~T and temperatures from $4$~K to $125$~K. The sample was mounted on an XYZ piezoelectric positioner in a closed-cycle cryostat equipped with a superconducting magnet, in static helium exchange gas. The sample temperature was measured with a calibrated sensor placed at the sample holder.

The Raman scattering signal was measured in the backscattering configuration using a confocal micro-Raman system with free-beam optics. The sample was excited with a continuous-wave laser at 633 nm, delivered to the microscope through a single-mode optical fiber and focused on the sample with a low-temperature apochromatic objective with a numerical aperture NA~=~0.81. The laser power was 1~mW, and each spectrum was recorded as 10 accumulations with an integration time of 60~s each. Measurements of the anti-Stokes spectral region showed no heating at the used laser power. The scattered light was collected through the same objective, filtered with a set of ultra-low-frequency filters, which allows the detection of Raman signals down to $\approx 10$~cm$^{-1}$, dispersed by a spectrometer with a focal length of 400~mm equipped with a 1200~grooves/mm grating, and detected with a thermoelectrically cooled back-illuminated deep-depletion CCD camera, which yields a spectral resolution of $\approx 0.5$~cm$^{-1}$ per pixel.

\section{EXPERIMENTAL RESULTS AND DISCUSSION}

\begin{figure}[t]
      \includegraphics[width=0.48\textwidth]{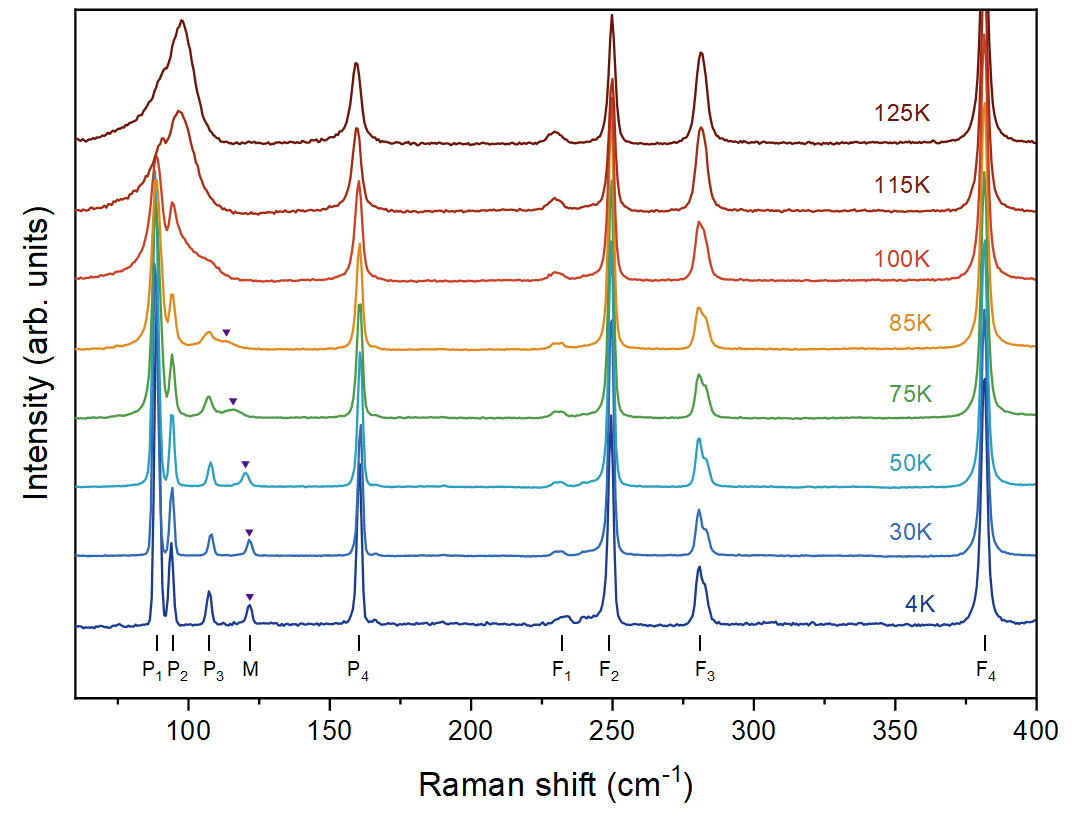}
      \caption{\label{Fig2} Raman spectra of FePS$_3$ measured at zero magnetic field at temperatures from 4 to $125$~K, stacked with a vertical offset. The bars mark the phonon modes $P_1$--$P_3$, $P_4$ and $F_1$--$F_4$ and the magnon $M$. Triangles mark the maximum of the magnon peak. The spectra are normalized to the intensity of the $F_4$ mode at $\approx 381$~cm$^{-1}$.}
\end{figure}

We begin the presentation of the experimental results by examining the temperature dependence of the Raman spectrum of FePS$_3$ in the absence of an applied magnetic field (Fig.~\ref{Fig2}). The results are in good agreement with previously reported data \cite{McCrearyPRB20, LeeNL16, ScagliottiPRB87, SekinePRB90}. The excitations labeled $P_i$ and $F_i$ correspond to phonon or phonon-like modes. The $P_i$ modes, involve displacements of the magnetic ions and are therefore sensitive to the magnetic order \cite{LeeNL16, BalkanskiJPC1987}. In particular, three well-resolved modes, $P_1$–$P_3$, are observed at low temperatures. These modes merge into a broad resonance above the Néel temperature, and are likely associated with zone-folded phonons arising from the doubling of the magnetic unit cell in the antiferromagnetic phase of FePS$_3$ \cite{LeeNL16, HashemiJPCC17}. In contrast, the higher-energy $F_i$ modes originate predominantly from vibrations of the (P$_2$S$_6$)$^{4-}$ anionic units and exhibit comparatively weak sensitivity to the magnetic order. The Raman scattering peak labeled as $M$, with its characteristic energy of $\approx 121$~cm$^{-1}$ in the spectra measured at $4$~K, is attributed to the magnon excitation \cite{McCrearyPRB20, Wyzula, SekinePRB90}. The $M$ mode exhibits a clear softening with increasing temperature, approaching the $P_3$ phonon mode and eventually merging with it at temperatures above $90$~K. Previous magneto-Raman scattering studies of antiferromagnetic FePS$_3$ performed at low temperatures revealed a pronounced coupling between the $M$ and $P_3$ modes \cite{Liu2021, PhysRevB.104.134437, Sun}. Despite the decreasing energy separation between the $M$ and $P_3$ modes upon heating, the characteristic energy of the $P_3$ phonon changes relatively little. This observation provides a preliminary indication that the magnon–phonon coupling may weaken with increasing temperature, an aspect that we address in greater detail in the following experiments. 

\begin{figure*}
     \includegraphics[width=1.0\textwidth]{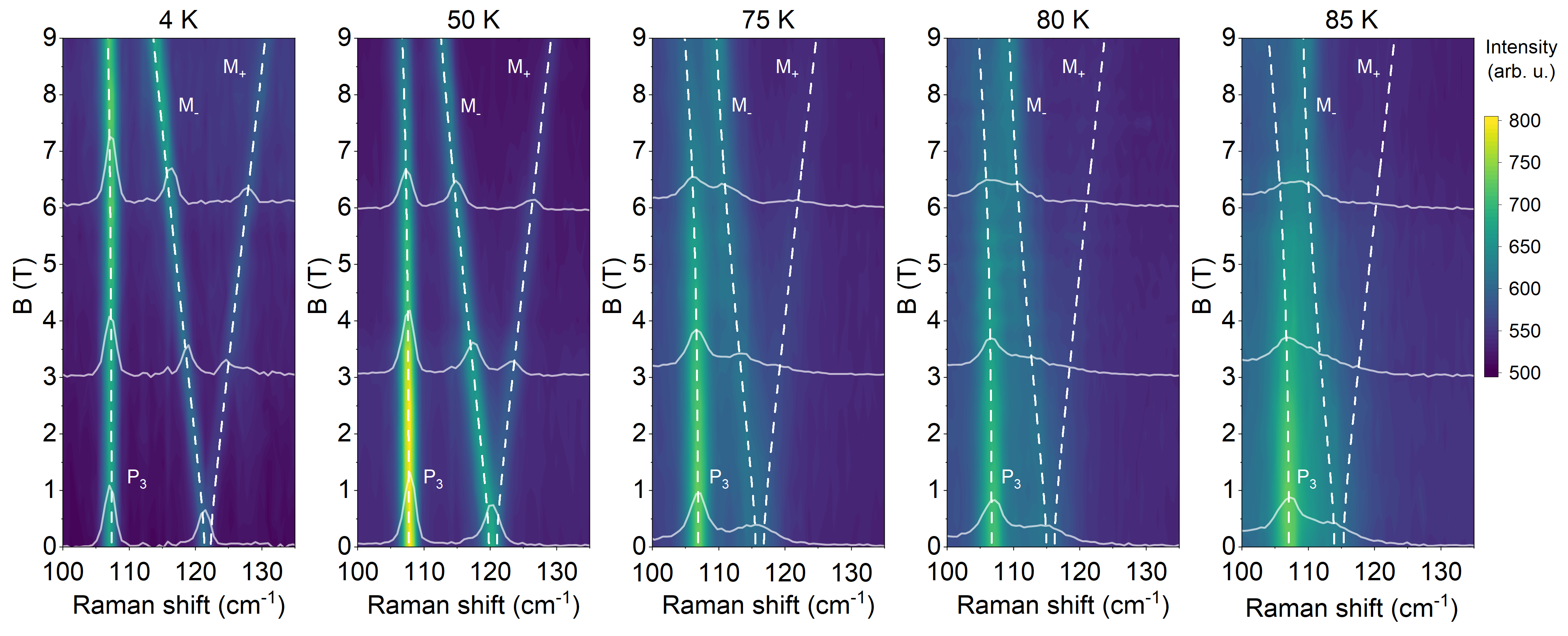}
     \caption{\label{Fig3} Color maps of magneto-Raman scattering spectra of FePS$_3$ measured at different temperatures: 4, 50, 75, 80 and $85$~K (color scale: Raman intensity). Dashed lines show the energies of the modes $M_+$, $M_-$ and $P_3$ calculated with the model described in the text, and the white lines show Raman spectra measured at selected magnetic fields.}
\end{figure*}

We now focus on the $M$ and $P_3$ modes and examine their evolution under an applied magnetic field at different temperatures. Figure~\ref{Fig3} presents the color maps of the magneto-Raman response along with the selective spectra measured in the range $100-135$~cm$^{-1}$ as a function of the applied magnetic field up to $9$~T, at different temperatures ranging from 4 to $85$~K. Following the antiferromagnetic resonance framework, the magnon peak splits into two Zeeman-like components, $M_-$ and $M_+$. As the applied magnetic field increases, the lower-energy magnon component, $M_-$, approaches the $P_3$ phonon mode. The interaction between these two modes results in the observed mode repulsion that ultimately leads to an avoided crossing at sufficiently high magnetic fields. Such anti-crossing point has previously been reported at 13.5~T for measurements performed at 4~K \cite{Liu2021, PhysRevB.104.134437}. In our low-temperature spectra, however, the avoided-crossing field lies beyond the accessible magnetic-field range. At higher temperatures ($T \geq 70$~K), the reduced energy separation between the $M$ and $P_3$ modes shifts the avoided crossing to lower fields, bringing it within the accessible range of our superconducting magnet. At temperatures above $85$~K, the magnon peak broadens and merges with the $P_3$ phonon, and the $M_-$ and $M_+$ components are no longer reliably distinguishable as a function of magnetic field. As the temperature increases further, the $P_1$ and $P_2$ bands broaden and begin to overlap with $P_3$ and $M$. From $110$~K, a single broad band dominates the spectrum. Magneto-Raman maps obtained in the 90 to $125$~K range are presented in the Supplemental Material.

To simulate the modes' energy patterns displayed in Fig.~\ref{Fig3}, we employ a phenomenological model accounting for the magnon–phonon interaction \cite{PhysRevB.104.134437, Pawbake}, in which we consider the two Zeeman-split magnon modes $M_+$ and $M_-$ and the phonon $P_3$, and assume the same coupling strength $\lambda$ between the phonon and each magnon branch. The Hamiltonian is $H=X^\dagger H' X$, with
$X = (\alpha_{M_+},\, \alpha_{M_-},\, \varsigma_{P_3},\, \alpha^{\dagger}_{M_+},\, \alpha^{\dagger}_{M_-},\, \varsigma^{\dagger}_{P_3})^{T}$,
where $\alpha$ and $\varsigma$ are the bosonic operators of the magnon and phonon modes, where:

\begin{align}
H^\prime &= \begin{pmatrix} F & G \\ G & F \end{pmatrix}, \quad
F = \begin{pmatrix} E_{M_+}^0 & 0 & \lambda \\ 0 & E_{M_-}^0 & \lambda \\ \lambda & \lambda & E_{P_3}^0 \end{pmatrix}, \nonumber \\
G &= \begin{pmatrix} 0 & 0 & \lambda \\ 0 & 0 & \lambda \\ \lambda & \lambda & 0 \end{pmatrix}, \label{eq:H}
\end{align}
\begin{equation}
  E_{M_\pm}^{0} = E_{M}^{0} \pm g\mu_\mathrm{B} B . \label{eq:EM}
\end{equation}

Here $E_{M}^{0}$ and $E_{P_3}^{0}$ are correspondingly the bare magnon and phonon energies at zero magnetic field; $g$ is the effective $g$-factor and $\mu_\mathrm{B}$ is the Bohr magneton. The Hamiltonian was diagonalized with the Bogoliubov transformation \cite{PhysRevB.99.064421, PhysRev.139.A450}, and the parameters $E_{M}^{0}$, $E_{P_3}^{0}$, $g$ and $\lambda$ were obtained for each temperature by a least-squares fit of the calculated energies to the measured ones. The simulations, shown with dashed lines in Fig.~\ref{Fig3}, reasonably reproduce the data at each temperature.

\begin{figure}
      \includegraphics[width=0.4\textwidth]{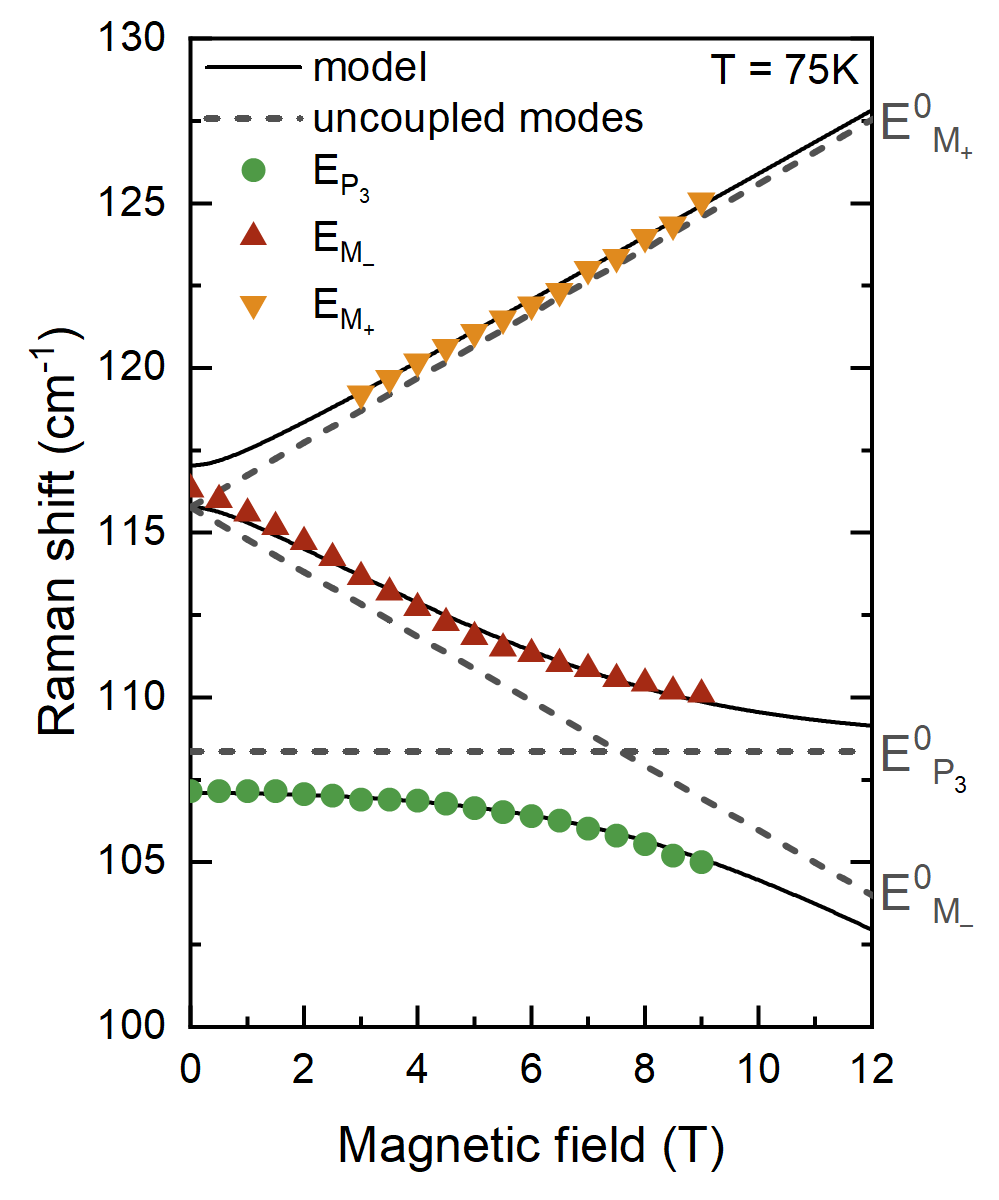}
      \caption{\label{Fig4} Magnetic-field dependence of the energies $E_{P_3}$, $E_{M_-}$ and $E_{M_+}$ of the hybrid magnon-phonon modes in FePS$_3$ at $T=75$~K, determined from magneto-Raman scattering spectra (symbols). Solid lines show results of simulations based on the model described in the text. Dashed lines indicate the field dependence of the energies of the bare (uncoupled) modes $E_{M_\pm}^0$ and $E_{P_3}^0$. Simulations are extended beyond the experimental range of $9$~T.}
\end{figure}

\begin{figure*}
     \includegraphics[width=0.8\textwidth]{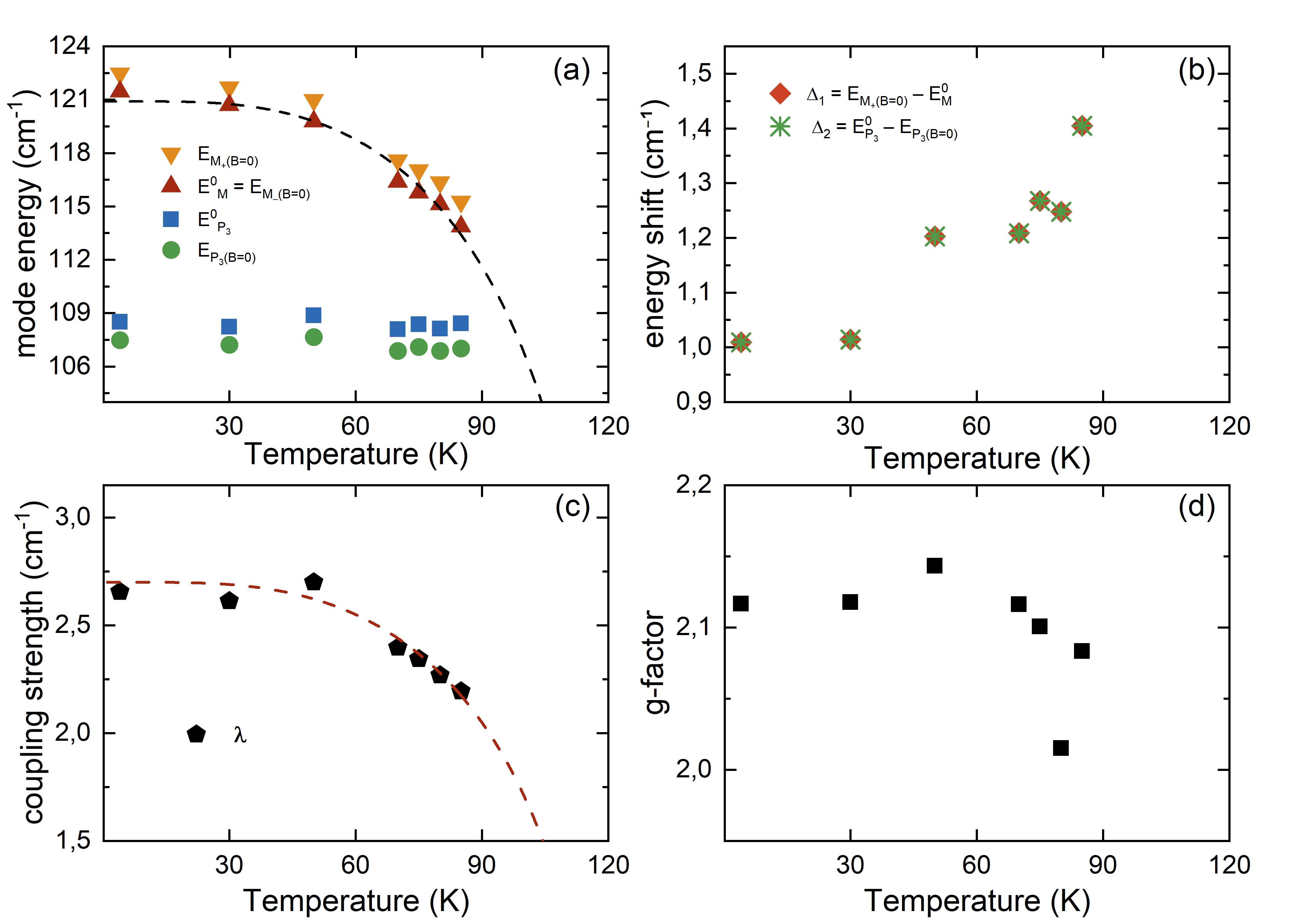}
     \caption{\label{Fig5} Temperature dependence of the parameters of the model. (a) Energies at $B=0$, extrapolated from the fits of the model: bare energies of the magnon $E_M^0$ and the phonon $E_{P_3}^0$ together with the zero-field energies of the hybrid modes $E_{M_+(B=0)}$ and $E_{P_3(B=0)}$; (b) zero-field shifts of the hybrid modes with respect to the bare energies, $\Delta_{1} = E_{M_{+}(B=0)} - E_{M}^{0}$ and $\Delta_{2} = E_{P_3}^{0}-E_{P_{3}(B=0)}$; (c) coupling constant $\lambda$; (d) $g$-factor. Dashed lines: Eq. (3) in (a) and (c).}
\end{figure*}

To illustrate the procedure in more detail, we focus on the data measured at $75$~K, at which the avoided crossing falls within the range of our magnet. Figure~\ref{Fig4} presents the experimentally determined energies $E_{P_3}$, $E_{M_-}$ and $E_{M_+}$ of the hybrid modes as a function of the magnetic field (symbols), together with the results of our data simulations. At $75$~K, the magnon peak is broadened, and the weak $M_+$ component is resolved only above $\approx3$~T. The simulated energies of the coupled magnon--phonon modes are shown as solid lines, and the energies of the bare (uncoupled) magnon modes, $E_{M_\pm}^0=E_M^0\pm g\mu_\mathrm{B}B$, and of the bare phonon mode, $E_{P_3}^0$, as dashed lines. The calculated traces are extended to $12$~T to show the evolution of the hybrid modes beyond the avoided crossing. As the magnetic field increases, the lower-energy magnon component, $M_-$, approaches the $P_3$ phonon mode, and the bare energies $E_{M_-}^0$ and $E_{P_3}^0$ cross at $B\approx7.6$~T. Near this field, the mode repulsion is strongest: the $P_3$ phonon mode is pushed below its bare energy and at higher fields follows the bare magnon energy $E_{M_-}^0$, whereas the $M_-$ mode levels off at $E_{P_3}^0$ and acquires a phonon-like character. The minimum separation between the two modes, $E_{M_-}-E_{P_3}\approx4.6$~cm$^{-1}$ at $\approx7.9$~T, is close to $2\lambda$. The avoided crossing thus provides a direct measure of the magnon--phonon coupling strength. In contrast, the upper magnon component, $M_+$, moves away from the $P_3$ phonon mode with increasing field and follows $E_{M_+}^0$, except for a small shift at $B=0$ discussed below. Repeating this procedure at each temperature yields the results shown in Fig.~\ref{Fig5}.

The effect of magnon--phonon coupling on the magnon mode at $B=0$ is a weak splitting into two, $M_{+}(B=0)$ and $M_{-}(B=0)$ components involving, correspondingly, the symmetric and antisymmetric combinations of the two magnon branches~\cite{PhysRevB.104.134437}. Since we assume that both magnon branches couple to the phonon with the same coupling strength $\lambda$, only the upper, $M_{+}(B=0)$, symmetric mode hybridizes with the $P_3$ phonon. In contrast, the lower, antisymmetric $M_{-}(B=0)$ mode corresponds to the bare magnon mode: $E_{M_{-}(B=0)}= E_{M}^{0}$. In our experiments, the two split magnon components at $B=0$ remain unresolved within the linewidth of the observed magnon mode (see Fig.~\ref{Fig3}). 
Meanwhile, the coupled $P_3$ phonon remains a single mode that is shifted to lower energy relative to its bare energy $E_{P_3}^0$.

The temperature evolution of the parameters derived from our data simulations is shown in Fig.~\ref{Fig5}. These parameters include the energy $E_{M}^{0}=E_{M_{-}(B=0)}$ of the bare magnon--mode, the energy $E_{P_3}^{0}$ of the bare phonon mode, the magnon--phonon coupling strength $\lambda$ (Fig.~\ref{Fig5}(c)), and the effective $g$-factor (Fig.~\ref{Fig5}(d)). This set of parameters is complemented in Fig.~\ref{Fig5}(a) by the energies $E_{M_{+}(B=0)}$ and $E_{P_{3}(B=0)}$ of the coupled magnon and phonon modes, respectively, at \textit{$B = 0$}. Because the $M_{+}(B=0)$ and $M_{-}(B=0)$ components are not resolved in our spectra, all energies shown in Fig.~\ref{Fig5}(a) are extrapolated to $B=0$ from the fits of the model to the experimentally determined magnetic field dependence of the energies of the $M_-$, $M_+$ and $P_3$ modes (see Fig.~\ref{Fig4}). The energy differences $\Delta_{1} = E_{M_{+}(B=0)} - E_{M}^{0}$ and $\Delta_{2} = E_{P_3}^{0}-E_{P_{3}(B=0)}$, which illustrate the mode repulsions, are also presented ({Fig.~\ref{Fig5}(b)).

Typically, for antiferromagnets, we observe that the bare magnon energy decreases with temperature. The $E_{M}^{0}(T)$ dependence is, however, surprisingly weak. It is difficult to account for this behavior using a conventional law, such as $E_{M}^{0}(T) = E_{M}^{0}{_{(T=0)}}(1-T/T_{N})^{\beta}$, for any reasonable value of the parameter $\beta$, when, as expected, we set $E_{M}^{0}{_{(T=0)}}$ = $121$~cm$^{-1}$ for the magnon energy in the limit of low temperature and $T_{N}$ = $118$~K for the N\'eel temperature. Nevertheless, in recent studies of the FePS$_3$ antiferromagnet \cite{Panda2026}, an optional BCS-type \cite{Carrington2003} formula:
\begin{equation}
  E_{M}^{0}(T) = a \tanh\!\left[ 1.82 \left( 1.018 \left( \frac{T_\mathrm{N}}{T} - 1 \right) \right)^{0.5} \right] + c ,
  \label{eq:tanh}
\end{equation}
has been proposed to account for temperature-induced softening of the magnon mode in this material. As shown by the dashed line in Fig.~\ref{Fig5}(a), the above formula reasonably reproduces our data points when we set $a=40.63$~cm$^{-1}$ and $c=80.26$~cm$^{-1}$ for the two fitting parameters.

As the temperature increases, the energy separation between the magnon and phonon modes decreases, so their coupling is expected to enhance mode repulsion and modify their respective energies. This enhancement of mode repulsion can be seen in our data, although, as already mentioned above, it appears to be a rather weak effect. Nonetheless, we can see that at temperatures above $50$~K, when the decrease of the separation between magnon and phonon modes is readily apparent (see Fig.~\ref{Fig5}(a)), the coupled phonon mode is pushed down and the coupled upper component of the magnon mode (which hybridized with the phonon mode) is pushed up in energy. This effect is better clarified in Fig.~\ref{Fig5}(b)) where energy differences $\Delta_{1}$ and $\Delta_{2}$ are also presented.

The relatively weak enhancement of mode repulsion with temperature points towards a weakening of the magnon--phonon coupling strength with temperature. As shown in Fig.~\ref{Fig5}(c), the coupling strength $\lambda$ extracted from our simulations indeed decreases with temperature. The $\lambda(T)$ dependence resembles that of previously discussed for $E_{M}^{0}(T)$, and, as shown by the dashed line in Fig.~\ref{Fig5}(c), can be reproduced using Eq.~(\ref{eq:tanh}) (though with different fitting parameters, $a=2.75$~cm$^{-1}$ and $c=-0.06$~cm$^{-1}$). 

The similarity between the observed temperature dependencies of $E_{M}^{0}(T)$ and $\lambda(T)$ suggests that the reduction in sublattice magnetization due to thermal fluctuations, which is known to be responsible for the softening of the magnon mode, also leads to a weakening of the magnon–phonon coupling. Moreover, we note that the observed decrease in $\lambda(T)/\lambda(T=0)$ is even more pronounced than that in $E_{M}^{0}(T)/E_{M}^{0}(T=0)$. Although the magnon–phonon coupling may be expected to depend on the ordered magnetic moment \cite{StreibPRB19, Vaclavkova_2020, Wang2023PCCP}, our experimental results call for a more detailed theoretical analysis of this effect. For completeness, we note that the derived $E_{P_3}^{0}$ and $g$-factor parameters are nearly temperature independent.

\section{CONCLUSION}

We investigated the van der Waals antiferromagnet FePS$_3$ using magneto-Raman scattering measurements over a temperature range from $4$~K to $125$~K, extending slightly above the Néel temperature. By resolving the coupled magnon–phonon excitations in spectra measured at temperatures up to $85$~K, we have traced their evolution with magnetic field and modelled the observed mode repulsion (anticrossing). We have discussed the temperature dependence of the extracted parameters, including the characteristic energy of the bare magnon and the magnon–phonon coupling strength.

Our main finding is that, similarly to the well-known temperature-induced softening of the magnon excitation, the magnon–phonon coupling strength decreases significantly with increasing temperature. This reduction may also be attributed to the decrease in sublattice magnetization caused by thermal fluctuations, although this interpretation requires confirmation through appropriate theoretical modelling. Nevertheless, since magnon–phonon coupling governs energy exchange between the spin and lattice systems, its temperature dependence should be taken into account when studying the magnetization dynamics of FePS$_3$, including the coherent terahertz driving of magnon polarons~\cite{Luo2025}. The reported temperature-induced changes in magnon–phonon hybridization should also be considered when investigating the temperature dependence of magnon–phonon coupling in other van der Waals antiferromagnets, such as FePSe$_3$ and CoPS$_3$ in which the magnon is close in energy to a phonon.\vspace{+6pt}

\section{ACKNOWLEDGMENTS}

We acknowledge the support of “Center for Terahertz Research and Applications (CENTERA2)” project (FENG.02.01-IP.05-T004/23) carried out within the “International Research Agendas” program of the Foundation for Polish Science co-financed by the European Union under European Funds for a Smart Economy Programme. Co-funded by the European Union (ERC TERAPLASM \#101053716). Views and opinions expressed are however those of the author(s) only and do not necessarily reflect those of the European Union or the European Research Council Executive Agency. Neither the European Union nor the granting authority can be held responsible for them. Supported also by "Center for Terahertz Research and Applications (CENTERA)” project carried out within the 'International Research Agendas' programme of the Foundation for Polish Science co-financed by the European Union under the European Regional Development Fund."

\bibliography{Krajewska-FePS3}

\end{document}